\documentclass{article}
\usepackage{spconf,amsmath,graphicx,hyperref}
\usepackage{amssymb}
\usepackage{booktabs}
\usepackage{multirow}
\usepackage{makecell}
\usepackage{xcolor}
\usepackage{colortbl}
\usepackage{threeparttable}

\title{Generalizable Speech Deepfake Detection via Information Bottleneck Enhanced Adversarial Alignment}
\name{Pu Huang$^{1}$, Shouguang Wang$^{1*}$, Siya Yao$^{1}$, Mengchu Zhou$^{1,2}$\thanks{* Corresponding author.}}
\address{\normalsize$^{1}$ School of Information and Electronic Engineering, Zhejiang Gongshang University, Hangzhou, China \\
         \normalsize$^{2}$ Department of Electrical and Computer Engineering, New Jersey Institute of Technology, Newark, USA}
\begin{document}
%\ninept
%
\maketitle
\begin{abstract}
Neural speech synthesis techniques have enabled highly realistic speech deepfakes, posing major security risks. Speech deepfake detection is challenging due to distribution shifts across spoofing methods and variability in speakers, channels, and recording conditions. We explore learning shared discriminative features as a path to robust detection and propose Information Bottleneck enhanced Confidence-Aware Adversarial Network (IB-CAAN). Confidence-guided adversarial alignment adaptively suppresses attack-specific artifacts without erasing discriminative cues, while the information bottleneck removes nuisance variability to preserve transferable features. Experiments on ASVspoof 2019/2021, ASVspoof 5, and In-the-Wild demonstrate that IB-CAAN consistently outperforms baseline and achieves state-of-the-art performance on many benchmarks.
\end{abstract}
\section{Introduction}
\label{sec:intro}

Recent advances in text-to-speech and voice conversion have greatly improved the realism of synthetic speech. Such speech deepfakes are increasingly indistinguishable from natural speech, raising serious risks of fraud, disinformation, and privacy violations. These concerns highlight the urgent need for reliable detection methods~\cite{li2025survey}.

Speech deepfake detection is an out-of-distribution generalization problem: detectors trained on limited spoofing methods must generalize to unseen ones at deployment~\cite{MullerInTheWild}. This mismatch across spoofing methods induces \textit{concept shift} (overfitting to attack-specific artifacts). In addition, \textit{covariate shift} arises from variability in speakers, codecs, and recording conditions, further degrading performance~\cite{cohen2022study}.

Existing strategies such as data augmentation~\cite{Tak2022Rawboost}, transfer learning~\cite{wang2021investigating, DBLP:conf/interspeech/Pan0SW24}, and representation optimization~\cite{DBLP:journals/tifs/XieCWY24, Zhang2021One-ClassLearning} improve robustness but often remain tied to attack-specific artifacts. A promising but under-explored direction is to learn discriminative cues shared across spoofing methods. This raises a natural question: do such shared discriminative features exist? Fortunately, empirical evidence indicates that they do indeed. For example,~\cite{DBLP:journals/corr/abs-2503-22503} demonstrates that different neural speech synthesis methods share common artifacts that can be captured by Transformer-based networks. Similarly,~\cite{wang2021investigating} shows that discriminative information within the 0.1–2.4 kHz band generalizes across datasets, while high-frequency cues may not transfer well. To leverage such cues, a common strategy is domain-invariant representation learning (DIRL)~\cite{wang2022generalizing}, such as domain adversarial training~\cite{DBLP:journals/jmlr/GaninUAGLLML16}. Yet in speech deepfake detection, the concept of ``domain'' is more subtle: artifacts function both as sources of domain shift and as critical discriminative signals. Consequently, blindly enforcing alignment may suppress informative cues while preserving irrelevant variability, thereby undermining generalization.

To address this, we propose Information Bottleneck enhanced Confidence-Aware Adversarial Network (IB-CAAN). CAAN leverages classifier confidence as an auxiliary adversarial signal to adaptively suppress harmful attack-specific artifacts, while the IB encourages compression of irrelevant input factors. Together, they guide the model to preserve cross-attack discriminative cues and learn minimal sufficient representations. Our contributions are summarized as follows:

1) We formalize speech deepfake detection as a dual distribution shift problem and propose to address it via attack-invariant feature learning.

2) We propose IB-CAAN, a novel attack-invariant feature learning framework for speech deepfake detection.

3) Experiments on ASVspoof 2019 LA~\cite{WangASVspoof2019}, ASVspoof 2021 LA and DF~\cite{Yamagishi2021ASVspoof2021}, ASVspoof 5~\cite{WangASVspoof5}, and In-the-Wild~\cite{MullerInTheWild} demonstrate that IB-CAAN outperforms baseline and surpasses state-of-the-art systems on many benchmarks. 

\begin{figure*}[t]
\centering
\includegraphics[width=0.8\linewidth]{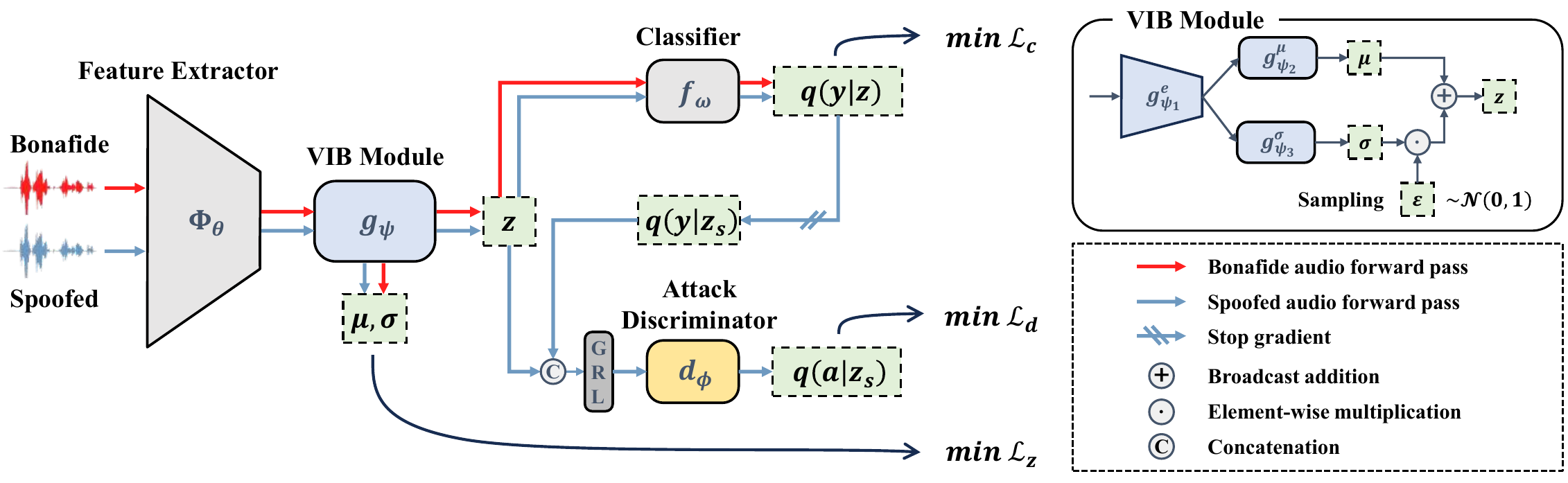}
\caption{The training pipeline of the proposed IB-CAAN. During inference, only $\Phi_\theta$, $g_\psi$, and $f_\omega$ are retained.}
\label{Framework}
\end{figure*}

\section{Methods}
\label{sec:format}

\subsection{Problem background}

In real-world speech deepfake detection tasks, the training phase typically relies on a limited dataset
\begin{equation}
D_{\text{tr}}=\{(x_i,y_i)\}_{i=1}^n,\;(x_i,y_i)\sim P_{\text{tr}}(X,Y),
\end{equation}
where $X,Y$ denote the input and label random variables with domains $\mathcal{X},\mathcal{Y} \,(\mathcal{Y}=\{0,1\})$, and $P_{\text{tr}}(X,Y)$ is joint distribution. However, the deployed model must work in a much more complex testing environment with a different joint distribution $P_{\text{te}}(X,Y)$. This mismatch manifests as two types of distribution shift: \textit{covariate shift} ($P_{\text{tr}}(X)\neq P_{\text{te}}(X)$, e.g., non-overlapping speakers) and \textit{concept shift} ($P_{\text{tr}}(Y|X)\neq P_{\text{te}}(Y|X)$, e.g., unseen spoofing algorithms breaking the model’s learned decision rule).

Our goal is to learn a binary classifier $h:\mathcal{X}\to\mathcal{Y}$, that minimizes the expected risk on the test distribution:
\begin{equation}
R(h)=\mathbb{E}_{(x,y)\sim P_{\text{te}}(X,Y)}[\ell(h(x),y)],
\end{equation}
where $\ell(\cdot,\cdot)$ denotes the $0$–$1$ loss or a differentiable surrogate. However, Empirical Risk Minimization (ERM) minimizes the empirical risk on $D_{\text{tr}}$, i.e.,
$\hat{R}_{\text{tr}}(h)=\tfrac{1}{n}\sum_{i=1}^n \ell(h(x_i),y_i),$
which generally fails to ensure the optimality of $R(h)$.

\subsection{Method overview}

To address these challenges, we propose IB-CAAN, as illustrated in Fig.~\ref{Framework}. In our design, CAAN encourages the learning of attack-invariant discriminative features, while the IB principle compresses redundant information.

% Formally, given an input speech sample $x$, we first employ a feature extractor to obtain the hidden representation h = f(x). We then use a variational information bottleneck to model the conditional distribution $p(z \mid x)$, yielding a compressed latent variable $z$ that retains the most relevant information for the spoofing detection task while discarding redundancy. Meanwhile, an attack-invariant representation learning constraint is applied to $z$ through adversarial training with a confidence-aware discriminator, encouraging the model to capture robust and discriminative features that are invariant to spoofing-specific artifacts.

\subsection{Information bottleneck}

To mitigate \textit{covariate shift}, we adopt the IB principle~\cite{DBLP:journals/corr/physics-0004057}. It seeks a latent representation $Z$ of input $X$ that compresses irrelevant information while retaining maximal predictive information about label $Y$. The IB objective is formulated as:
\begin{equation}
\min \; \beta I(X; Z) - I(Z; Y),    \label{eq:3}
\end{equation}
where $I(X; Z)$ denotes the mutual information between input $X$ and latent representation $Z$, and $I(Z; Y)$ denotes the mutual information between $Z$ and output $Y$. $\beta$ controls the trade-off between compression and prediction. Since mutual information in high-dimensional continuous spaces is intractable to compute directly, optimizing Eq.~(\ref{eq:3}) is generally infeasible. We follow the Variational Information Bottleneck (VIB)~\cite{DBLP:conf/iclr/AlemiFD017} and optimize its variational upper bound:
\begin{align}
\beta I(X; Z) - I(Z; Y) &\leq \beta \, \mathbb{E}_{p(x,z)}\left[\log \left( \frac{p(z \mid x)}{r(z)} \right)\right] \nonumber \\
\quad &-\mathbb{E}_{p(x,y,z)}[\log q(y \mid z)], \label{eq:4}
\end{align}
where $r(z)$ and $q(y\mid z)$ are variational approximations to the true marginal $p(z)$ and conditional $p(y\mid z)$, respectively. $p(z\mid x)$ is the conditional distribution given by a encoder. Since sampling directly from $z \sim p(z\mid x)$ prevents backpropagation, we apply the reparameterization trick, which rewrites the sampling process as a deterministic function of an auxiliary noise variable $\epsilon$, i.e., $z = \mu(x) + \sigma(x) \odot \epsilon, \quad \epsilon \sim \mathcal{N}(0, I).$
We assume $r(z) = \mathcal{N}(0, I)$. The encoder is parameterized by a feature extractor $\Phi_\theta(\cdot)$ and an VIB module $g_\psi(\cdot)$ which include an MLP encoder $g_{\psi_1}^e(\cdot)$, and two linear projections $g_{\psi_2}^{\mu}(\cdot)$ and $g_{\psi_3}^{\sigma}(\cdot)$. Thus, the conditional distribution is modeled as $p(z \mid x;\, g \circ \Phi) = \mathcal{N}\!\big(\mu(x), \text{diag}(\sigma(x)^2)\big)$, with $\mu(x) = g_{\psi_2}^{\mu}(g_{\psi_1}^e(\Phi_\theta(x))), \; \sigma(x) = g_{\psi_3}^{\sigma}(g_{\psi_1}^e(\Phi_\theta(x)))$.
Finally, we optimize the following training objective:
\begin{align}
\mathcal{L}_{\text{c}} + \beta\mathcal{L}_{\text{z}} = \min_{\theta,\psi,\omega} \; &\mathbb{E}_{z \sim p(z \mid x;\, g \circ \Phi), y}\big[\mathcal{L_{\text{wce}}}(f_\omega(z), y)\big] \; \nonumber \\
+\; \beta \min_{\theta,\psi}\; &\mathbb{E}_{x} \Big[ \mathrm{KL}\big(p(z \mid x;\, g \circ \Phi) \,\|\, r(z)\big)\Big]. \label{eq:6}
\end{align}
Where $f_\omega$ is a binary classifier, $\mathcal{L}_{\text{wce}}$ is weighted cross-entropy loss, and the Kullback-Leibler (KL) divergence regularizes the latent representation.

\subsection{Confidence-aware adversarial Network}

To alleviate \textit{concept shift}, our goal is to enable the model to learn generalizable discriminative features shared across different spoofing algorithms. In domain generalization~\cite{wang2022generalizing}, this idea is formalized as DIRL. Let $z_s$ denote the latent representation of spoofed speech obtained from the VIB encoder, i.e.,
$z_s \sim p(z \mid x_s;\, g \circ \Phi)$, $x_s$ are spoofed samples.
And a main classifier $f_\omega(\cdot)$ predicts the binary label (bonafide or spoof), while a domain discriminator $d_\phi(\cdot)$ distinguishes spoof types. Following  DANN~\cite{DBLP:journals/jmlr/GaninUAGLLML16}, we adopt an adversarial objective between VIB module $g_\psi$ and discriminator $d_\phi$:
\begin{align}
\mathcal{L}_\text{c} + \alpha\mathcal{L}_{\text{d}} = &\min_{\theta,\psi,\omega} \;\mathbb{E}_{z_s, y}\mathcal{L}_{\text{wce}}\big(f_\omega(z_s), y\big) \nonumber \\
 +\; \alpha &\,\min_{\phi} \max_{\theta,\psi}\;\mathbb{E}_{z_s, a}\mathcal{L}_{\text{ce}}\big(d_\phi(z_s), a\big), \label{eq:8}
\end{align}
where $\mathcal{L}_\text{c}$ is the same main classification objective as defined in Eq.~(\ref{eq:6}), $\mathcal{L}_{\text{ce}}$ is cross-entropy loss, and $a$ is spoof type label. The min–max structure encourages the latent representation $z_s$ to be invariant to spoof types. In practice, the maximization over $(\theta,\psi)$ is implemented via a gradient reversal layer (GRL), which keeps the forward pass unchanged but multiplies a gradient by $-\lambda$ during backward.

In the context of \textit{concept shift}, the ``domain'' is induced by spoof-specific artifacts that are correlated with the main task. To analyze this, we decompose the speech feature as $X\triangleq X_0\oplus X_1\oplus X_2$, where $X_0$ denotes task-irrelevant variations, $X_1$ represents attack-invariant discriminative features, and $X_2$ captures attack-specific characteristics. However, although standard DANN aims to suppress $X_2$, it may inadvertently amplify $X_0$, thereby conflicting with the main task.

To address this issue, we propose the CAAN framework, in which the discriminator receives as input both the latent representations and the classifier’s confidence scores. Accordingly, the second term in Eq.~(\ref{eq:8}) is reformulated as
\begin{equation}
\alpha\mathcal{L}_{\text{d}} = \alpha\,\min_{\phi} \max_{\theta,\psi}\;\mathbb{E}_{z_s, a}\mathcal{L}_{\text{ce}}\big(d_\phi([z_s, c(z_s)]), a\big), \label{eq:9}
\end{equation}
where $c(z_s) = 1/(1+e^{-f_\omega(z_s)})$ denotes main classifier confidences, $[\cdot]$ denotes the concatenation operation. With this design, when classifier confidence is high, $\mathcal{L}_{\text{d}}$ encourages alignment that suppresses attack-specific artifacts; when confidence is low, it instead reduces spurious alignment.

\subsection{Training objective}

Finally, we combine the IB with CAAN by jointly optimizing:
\begin{equation}
\mathcal{L} = \mathcal{L}_{c} + \beta\mathcal{L}_{z} + \alpha\mathcal{L}_{d},
\end{equation}
where $\mathcal{L}_{c}$ is binary classification loss, $\mathcal{L}_{z}$ is IB regularization term, and $\mathcal{L}_{d}$ is adversarial loss. $\beta$ and $\alpha$ are hyperparameters that balance the loss terms.

\section{EXPERIMENTS}
\label{sec:pagestyle}

\subsection{Experimental setups}

{\bf Datasets and metrics.}
To evaluate cross-dataset generalization, we follow the official protocols of ASVspoof 2019~\cite{WangASVspoof2019}, ASVspoof 2021~\cite{Yamagishi2021ASVspoof2021}, and ASVspoof 5~\cite{WangASVspoof5}. Two groups of experiments are conducted: 1) Training on the ASVspoof 2019 LA training set (ASV19:Trn) and evaluating on eval sets of ASVspoof 2019 LA (19LA), ASVspoof 2021 LA (21LA) and ASVspoof 2021 DF (21DF); 2) Training on the ASVspoof 5 training set (ASV5:Trn) and evaluating on ASVspoof 5 Track 1 eval set (ASV5:T1). Additionally, we perform evaluations on In-the-Wild (ITW)~\cite{MullerInTheWild} under both training protocols.
% The ASVspoof 2019 and 2021 corpora are derived from VCTK, featuring controlled recording conditions but progressively more diverse spoofing techniques (19LA with 13 TTS/VC systems, 21LA with additional codec/channel variability, and 21DF with over 100 spoofing algorithms). ASVspoof 5, based on MLS English, covers thousands of speakers with latest TTS/VC algorithms. The ITW dataset, collected from online deepfake videos, reflects unconstrained real-world conditions.
We report performance using Equal Error Rate (EER)~\cite{li2025survey}.

{\bf Implementation details.}
All audio samples used for training and evaluation are cropped or zero-padded to approximately 4 seconds (64,600 samples). We adopt three representative backbones, which are RawBMamba~\cite{DBLP:conf/interspeech/ChenYXW0DZ0LF24}, XLSR+Linear~\cite{wang2021investigating}, and XLSR+MLP~\cite{DBLP:conf/icassp/WangY23}, with different training setups. Specifically, RawBMamba uses batch size $32$, learning rate $1\times10^{-5}$, weight decay $1\times10^{-4}$, Adam optimizer~\cite{Kingma2015Adam} with $\beta_1=0.9,~\beta_2=0.98$, and Noam scheduler~\cite{DBLP:conf/nips/VaswaniSPUJGKP17}, while XLSR-based backbones use batch size 12, learning rate $1\times10^{-6}$, weight decay $1\times10^{-4}$, Adam optimizer with $\beta_1=0.9,\beta_2=0.999$, and no scheduler. For the trade-off coefficients in training loss, we set $(\beta,~\alpha)$ to $(0.001,1)$ for RawBMamba on ASV19:Trn, $(0.001,0.5)$ for XLSR-based backbones on ASV19:Trn, $(0.001,0.5)$ for RawBMamba on ASV5:Trn, and $(0.0 1,0.5)$ for XLSR-based backbones on ASV5:Trn. For the GRL coefficient $\lambda$, we follow~\cite{DBLP:journals/jmlr/GaninUAGLLML16} and adopt a schedule $\lambda_p = \frac{2}{1+\exp(-10\cdot p)} - 1$, which increases smoothly from 0 to 1. We consider both settings with and without data augmentation. In the augmentation setting, RawBoost~\cite{Tak2022Rawboost} is applied to ASV19:Trn, and MUSAN noise~\cite{DBLP:journals/corr/SnyderCP15} together with RIR~\cite{DBLP:conf/icassp/KoPPSK17} are applied to ASV5:Trn. Final results were obtained by averaging the top-5 checkpoints from 30 training epochs\footnote{Code available at~\url{https://github.com/763021701/IB-CAAN}}.

\begin{table*}[t]
    \caption{Comparison with baselines in EER (\%). The results are reported as the format of \texttt{best/mean} across 3 runs.}
    \label{Comparison Baseline}
    \setlength{\tabcolsep}{3.8pt}
    \small
    \centering
    \begin{tabular}{ccccccccccc}
      \toprule
      \multirow{2}{*}{$\Phi_\theta$} & \multirow{2}{*}{$f_\omega$}  & \multirow{2}{*}{Algorithm}  & \multicolumn{4}{c}{Train on ASV19:Trn}   & \multicolumn{3}{c}{Train on ASV5:Trn}\\
      \cmidrule(lr){4-7}
      \cmidrule(lr){8-10}
      & & & 19LA & 21LA & 21DF & ITW & ASV5:Dev & ASV5:T1 & ITW \\
      % \midrule
      % \multirow{2}{*}{RawNet2} & \multirow{2}{*}{Linear}
      %                            & ERM & 4.76 / 5.62 & 8.85 / 9.48 & 22.06 / 22.27 & 46.76 / 49.39 & 31.85 / 32.44 & 38.06 / 38.48 & 22.14 / 23.42 \\
      %                          & & IB-CAAN  & \textbf{4.29} & \textbf{8.63} & \textbf{22.03} & \textbf{48.28} & \textbf{30.67} / \textbf{31.81} & \textbf{35.46} / \textbf{37.04} & \textbf{17.11} / \textbf{19.51} \\
      \midrule
      \multirow{2}{*}{RawBMamba} & \multirow{2}{*}{Linear}
                                 & ERM & 1.78 / \textbf{1.91} & 5.22 / 5.46 & \textbf{19.42} / 20.09 & 42.18 / 42.62 & 20.55 / 21.80 & 31.19 / 32.77 & 30.77 / 35.17  \\
                               & & IB-CAAN  & \textbf{1.71} / 2.24 & \textbf{5.01} / \textbf{5.43} & 19.43 / \textbf{19.90} & \textbf{24.60} / \textbf{27.85} & \textbf{17.96} / \textbf{18.52} & \textbf{30.59} / \textbf{31.01} & \textbf{21.22} / \textbf{24.46} \\
      \midrule
      \multirow{2}{*}{XLSR} & \multirow{2}{*}{Linear} 
                                 & ERM & 0.53 / \textbf{0.57} & 4.89 / 6.70 & 4.41 / 5.61 & 10.15 / 14.13 & 0.33 / 0.43 & 8.17 / 8.38 & 17.34 / 19.22 \\
                               & & IB-CAAN  & \textbf{0.37} / 0.58 & \textbf{4.66} / \textbf{5.06} & \textbf{3.28} / \textbf{3.51} & \textbf{5.54} / \textbf{5.99} & \textbf{0.18} / \textbf{0.26} & \textbf{6.26} / \textbf{6.31} & \textbf{8.51} / \textbf{13.48}  \\
      \midrule
      \multirow{2}{*}{XLSR} & \multirow{2}{*}{MLP}
                                 & ERM & 0.56 / 0.80  & 5.16 / 7.55  &  6.12 / 7.40  & 9.41 / 10.74 & 0.41 / 0.53 & 6.85 / 8.79 & 17.15 / 18.99 \\
                               & & IB-CAAN  & \textbf{0.24} / \textbf{0.40} & \textbf{4.00} / \textbf{4.69} & \textbf{3.50} / \textbf{3.75} & \textbf{4.61} / \textbf{4.93} & \textbf{0.09} / \textbf{0.12} & \textbf{5.96} / \textbf{5.98} & \textbf{8.11} / \textbf{11.62} \\
      \bottomrule
    \end{tabular}
\end{table*}

% \begin{table}[t]
%     \caption{Training setups for different backbones. 
%     Trn = Training set, BS = Batch size, 
%     LR = Learning rate, WD = Weight decay, Sch. = LR scheduler.}
%     \label{training detail}
%     \centering
%     \small
%     \begin{tabular}{l|ccccc}
%       \toprule
%       \textbf{Backbone}  & \textbf{BS} & \textbf{LR} & \textbf{WD} & \textbf{Adam} & \textbf{Sch.} \\
%       \midrule
%       RawNet2 & 32  & $1 \times 10^{-4}$ & $1 \times 10^{-4}$ & [0.9, 0.999] & None \\
%       RawBMamba  & 32 & $1 \times 10^{-5}$ & $1 \times 10^{-4}$ & [0.9, 0.98] & Noam \\
%       \midrule
%       \makecell[l]{XLSR-Linear \&\\ XLSR-MLP} & 12 & $1 \times 10^{-6}$ & $1 \times 10^{-4}$ & [0.9, 0.999] & None \\
%       \bottomrule
%     \end{tabular}
% \end{table}

\begin{table}[t]
    \caption{Results of comparison with state-of-the-art single systems on 19LA, 21LA, 21DF, and ITW in EER (\%). Best results are in \textbf{bold}, and second-best results are \underline{underlined}.}
    \label{Comparison SOTA:1}
    \setlength{\tabcolsep}{4pt}
    \small
    \centering
    \begin{threeparttable}
    \begin{tabular}{lcccc}
      \toprule
      System  & 19LA  & 21LA  & 21DF  &  ITW  \\
      \midrule
      2022 XLSR+AASIST~\cite{Tak2022AutomaticSpeakerVerificationSpoofing} & -    & 6.15 & 7.70 & - \\
      2022 XLSR+AASIST*~\cite{DBLP:conf/icassp/LuZWSZ24} & \underline{0.22} & \textbf{0.82} & 2.85 & 10.48 \\
      % 2023 XLSR+Linear* [xx] & 0.22 & 3.63 & 3.65 & 16.17 \\
      2024 OCKD*~\cite{DBLP:conf/icassp/LuZWSZ24}        & 0.39 & \underline{0.90} & 2.27 & 7.68  \\
      2025 XLSR+MoE*~\cite{DBLP:conf/icassp/WangFW0WXQSLLLL25}    & -    & 2.96 & 2.54 & 9.17  \\
      2024 XLSR+SLS*~\cite{DBLP:conf/mm/ZhangWH24}    & -    & 3.88 & 2.09 & 8.87  \\
      2025 LSR+LSA*~\cite{DBLP:conf/icassp/0004GWZQ25}    & \textbf{0.15} & 1.19 & 2.43 & 5.92  \\
      2025 XLSR+Mamba*~\cite{DBLP:journals/spl/XiaoD25}  & -    &  0.93 & 1.88 & 6.71  \\
      2025 XLSR+Nes2Net-X*~\cite{DBLP:journals/corr/abs-2504-05657} & -    & 2.00 &  \underline{1.78} & 6.60  \\
      \midrule
      XLSR+MLP (IB-CAAN)  & 0.40 & 4.69 & 3.75 & \textbf{4.93}  \\
      XLSR+MLP (IB-CAAN)*    & 0.31 & 2.21  & \textbf{1.64} &  \underline{5.65} \\
      \bottomrule
    \end{tabular}
    \begin{tablenotes}
    \footnotesize
    \item * with Rawboost augmentation.
    \end{tablenotes}
    \end{threeparttable}
\end{table}

\begin{table}[t]
    \caption{Results of comparison with state-of-the-art single systems on ASV5:T1 in EER (\%). Best results are in \textbf{bold}.}
    \label{Comparison SOTA:2}
    \setlength{\tabcolsep}{4pt}
    \small
    \centering
    \begin{threeparttable}
    \begin{tabular}{clc}
      \toprule
      Condition  & System  & EER  \\
      \midrule
      \multirow{5}{*}{Closed} & 2024 T06~\cite{WangASVspoof5}  & 28.41  \\
                              & 2024 T19~\cite{WangASVspoof5}  & 24.59  \\
                              & 2024 T48~\cite{WangASVspoof5}  & \textbf{23.63}  \\
      \cmidrule(lr){2-3}
                              & RawBMamba (IB-CAAN)  & 31.01 \\
                              & RawBMamba (IB-CAAN)*  & 27.77 \\
                
      \midrule
      \multirow{7}{*}{Open}   & 2022 XLSR+AASIST*~\cite{DBLP:journals/corr/abs-2504-05657}  & 6.08 \\
                              & 2024 T19~\cite{WangASVspoof5}  & 6.06 \\
                              & 2024 T17~\cite{WangASVspoof5}  & 5.99 \\
                              & 2025 XLSR+Nes2Net-X*~\cite{DBLP:journals/corr/abs-2504-05657}  & 5.92 \\
                              & 2024 T31~\cite{WangASVspoof5}  & 5.56 \\
      \cmidrule(lr){2-3}
                              & XLSR+MLP (IB-CAAN) & 5.98 \\
                              & XLSR+MLP (IB-CAAN)* &  \textbf{4.67} \\
      \bottomrule
    \end{tabular}
    \begin{tablenotes}
    \footnotesize
    \item * with Musan and RIR augmentations.
    \end{tablenotes}
    \end{threeparttable}
\end{table}

\begin{table}[t]
    \caption{Results of ablation experiments. Best results are in \textbf{bold}, and second-best results are \underline{underlined}.}
    \label{Ablation}
    \setlength{\tabcolsep}{3.5pt}
    \small
    \centering
    \begin{tabular}{lcccccc}
      \toprule
      Configuration   & 19LA  & 21LA   &  21DF & ITW & ASV5:T1 & AVG   \\
      \midrule
      IB-CAAN  & \textbf{0.40} & \underline{4.69} & \textbf{3.75} & \textbf{4.93} & \textbf{5.98} & \textbf{3.95} \\
      \quad  w/o IB  & 1.07 & 10.33 & 18.29 & 23.40 & 16.50 & 13.92 \\
      \quad  w/o CAAN  & \underline{0.57} & 5.47 & \underline{4.45} & \underline{8.78} & 9.21 & 5.56 \\
      IB-DANN  & 1.03 & \textbf{4.00} & 4.72 & 10.26 & \underline{6.64} & \underline{5.33} \\
      \bottomrule
    \end{tabular}
\end{table}

\subsection{Comparison with baselines}

To evaluate the effectiveness of IB-CAAN, we compare it with ERM (baseline) using three backbones, as shown in Table~\ref{Comparison Baseline}. In in-domain evaluation (19LA), IB-CAAN achieves comparable performance to ERM with RawBMamba and XLSR+Linear, while with XLSR+MLP it reduces the EER from 0.80\% to 0.40\%, achieving a 50\% relative improvement. In out-of-domain evaluations (21LA, 21DF, ASV5:T1, and ITW), IB-CAAN consistently outperforms ERM, e.g., on ASV5:T1, IB-CAAN with XLSR+MLP reduces the EER from 8.79\% to 5.98\%, while on ITW it lowers the EER from 10.74\% to 4.93\%, showing consistent gains in cross-dataset evaluations. Notably, the gains are particularly large with XLSR-based backbones, which we attribute to the Transformer networks being more effective in capturing cross-attack discriminative cues, consistent with prior findings~\cite{DBLP:journals/corr/abs-2503-22503}.

\subsection{Comparison with state-of-the-art single systems}

As shown in Table~\ref{Comparison SOTA:1}, without data augmentation, XLSR+MLP (IB-CAAN) significantly outperforms XLSR+AASIST under the same setting, achieving the lowest EER of 4.93\% on ITW. With data augmentation, the EER further decreases to 1.64\% on 21DF, achieving the best performance, and our system still obtains the lowest EER on ITW with 5.65\%. On 21LA, the 2.21\% EER is comparable to recent advanced XLSR-based systems such as XLSR+MoE and XLSR+Nes2Net-X. For ASVspoof5 (Table~\ref{Comparison SOTA:2}), RawBMamba (IB-CAAN) achieves 27.77\% EER under the closed condition, falling short of the best-performing submission at 23.63\%, while remaining competitive compared to other strong systems. In contrast, under the open condition, XLSR+MLP (IB-CAAN) achieves 4.67\% EER, establishing a new state of the art.

\subsection{Ablation studies}

Table \ref{Ablation} presents the results of ablation experiments. Removing the IB module leads to a large degradation, with the average EER rising from 3.95\% to 13.92\%, showing that IB plays a key role in suppressing irrelevant variations. Excluding CAAN also worsens performance, especially on ITW and ASV5:T1. Removing confidence guidance (IB-DANN) improves performance on 21LA but causes clear degradation on other datasets. The complete IB-CAAN achieves the best overall (average) performance, demonstrating that IB and CAAN are complementary for improving generalization.

\section{CONCLUSIONS}
\label{conclusions}

In this work, we introduced IB-CAAN for speech deepfake detection. IB-CAAN compresses nuisance variability and suppresses attack-specific artifacts, which facilitates learning transferable discriminative features and improves generalization. Experiments on ASVspoof 2019/2021, ASVspoof 5, and In-the-Wild datasets demonstrate that IB-CAAN consistently outperforms the ERM baseline and achieves the best performance on many benchmarks. In future work, we will further investigate attack-invariant discriminative feature learning for speech deepfake detection.

% -------------------------------------------------------------------------
\vfill\pagebreak
\small
\section{ACKNOWLEDGMENT}
The authors used a large language model (LLM) to polish language and improve clarity. All content was reviewed, edited, and verified by the authors.
% -------------------------------------------------------------------------
\small
\bibliographystyle{IEEEbib}
\bibliography{references}

\end{document}